\documentclass[12pt,a4paper]{article}

\usepackage{a4wide}
\usepackage{ncmacros}

\newcommand{\be}{\begin{equation}}
\newcommand{\ee}{\end{equation}}
\newcommand{\ba}{\begin{eqnarray}}
\newcommand{\ea}{\end{eqnarray}}
\newcommand{\nn}{\nonumber}
\newcommand{\CI}{{\cal I}}
\begin{document}

%
%

\begin{flushright}
\textsc{thu-98/32\\ cern-th/98-270 \\ hep-th/9808128\\ August 1998}
\end{flushright}

\begin{center}

{\scshape\LARGE U-Duality Invariance of the Four-dimensional Born-Infeld 
Theory\par} \vskip4em

\textsc{Christiaan Hofman$^1$, Erik Verlinde$^{1,2}$ \textnormal{and} 
Gysbert Zwart$^1$}\\[2mm]

\textit{${}^1$Institute for Theoretical Physics, University of Utrecht\\
        Princetonplein 5, 3508 TA Utrecht}\\[1mm]
\textit{${}^2$TH-Division, CERN 
CH-1211 Geneva 23}\\[1mm]
\texttt{hofman@phys.uu.nl, verlinde@phys.uu.nl, zwart@phys.uu.nl}\\[5mm]
 
\end{center}
\vskip4em

\begin{abstract}

We calculate the Hamiltonian of a compactified D4-brane, with general 
fluxes and moduli, and find the BPS-mass. The results are 
invariant under the complete U-duality $\SO(5,5,\Z)$. 

\end{abstract}

%
%

\section{Introduction}

Toroidally compactified string theory is symmetric under a discrete group 
of U-dualities. These U-dualities act on the various charges present in 
the theory. In six dimensions, the U-duality group is $\SO(5,5,\Z)$; it 
includes an $\SO(4,4,\Z)$-part which is the T-duality group of string 
theory on a four-torus.
The enhancement to the full U-duality group is 
perhaps better understood from the eleven dimensional point of view; 
indeed, in \cite{dvv} the $\SO(5,5,\Z)$ symmetry was interpreted as a 
T-duality of strings living on the M-theory five-brane. 

From the ten-dimensional string point of view, the various charges can be 
split up in NSNS, perturbative ones (momenta and winding) and 
non-perturbative (RR) charges, carried by D0,2 and 4-branes wrapping 
cycles of the four-torus. The $\SO(4,4,\Z)$ subgroup respects this 
split-up, the full U-duality group however mixes both sectors. 

The compactified string theory is related to a $4+1$ dimensional gauge 
theory. The gauge theory is the world volume theory of the D4-branes: 
a stack of $N$ D4-branes gives rise to a $\U(N)$ gauge theory. Also the 
other charges are represented in the gauge theory, as electric and 
magnetic fluxes, compact momenta and the instanton number. Hence a 
given, generic set of string theory charges can be mapped to a gauge 
theory configuration with the corresponding set of quantum numbers.

As is well known, the correct action to describe the gauge theory is 
the Born-Infeld theory \cite{leigh}. Using this action, we calculate the 
Hamiltonian 
as a function of the various fields and moduli. Remarkably, we find a 
completely $\SO(5,5,\Z)$-invariant expression. We also derive a lower 
bound on the mass given the quantum numbers, the BPS-mass. This mass is 
identical to the one found in \cite{dvv} from a six-dimensional 
space-time supersymmetry calculation.
 
\section{Dual Pictures}

We are interested in IIA string theory 
compactified on a four-dimensional torus ${\bf T}^4$. On this torus we 
allow for a general configuration of D0-branes, wrapped 
D4- and D2-branes, winding strings and compact momenta. We 
want to identify the various conserved charges in this theory and derive 
the BPS-spectrum, as a function of the moduli. In the end we will 
demonstrate the invariance of 
this spectrum under the full U-duality group in six dimensions, 
$\SO(5,5,\Z)$. The system can be looked at from various points of 
view. The interpretation of both the charges and the moduli depends on 
the perspective. 
In the eleven-dimensional point of view the $\SO(5)$ 
symmetry of the U-duality group is manifest. The zero-brane number is in 
this picture interpreted as momentum around an additional circle. The 
quantum numbers are associated to five momenta, $10$ membrane-wrapping 
numbers and the five-brane number. The moduli are unified in a five 
dimensional metric ${G}$ and a 
three-form field ${C}_3$.
Together, ${G}$ and the 
Hodge dual of ${C}$ parametrise in the usual fashion the moduli 
space, 
\be
\CM_{5,5} =\frac{\SO(5,5)}{\SO(5)\times\SO(5)}.
\ee

A dual description is obtained via reduction on $S_1$ to get to 
the string picture. The M-theory five-branes are mapped to D4-branes, 
which are described by a non-abelian
$4+1$-dimensional Born-Infeld theory. The quantum numbers in 
this case are the rank of the gauge group, the electric and magnetic
fluxes on the four torus, the compact momenta of the gauge theory and the 
instanton number. The Born-Infeld action contains couplings to the
background metric $g$, $b$-field and RR-forms $c_1,c_3$, and is
multiplied by the string coupling $\lambda$.

The relations between the quantum numbers in the various descriptions of 
the theory are described in table \ref{tb:rel}. The sixteen charges 
transform under the U-duality group $\SO(5,5)$ in the sixteen-dimensional 
spinor representation.

\begin{table}
\begin{center}
\begin{tabular}{|c|c|c|c|}
\hline
quantum \# & strings & $(4+1)$d Born-Infeld & M-theory \\ \hline
$k$       & D0 & instanton number & $P_5$\\
$n_i$     & momentum & momentum & $P_i$ \\
$w_i$     & winding  & electric flux & $M_{5i}$\\
$m_{ij}$  & D2 & (dual of) magnetic flux & $M_{ij}$\\
$N$       & D4 & rank & five-brane\\
\hline
\end{tabular}
\end{center}
\caption{charges in the various pictures}
\label{tb:rel}
\end{table}

In the following we will derive the BPS-mass from the Born-Infeld action 
(the D4-branes point of view). We will restrict ourselves to the abelian 
case, i.e.  $N=1$. The resulting expression will then be 
expressed in terms of the M-theory variables to make its U-duality 
invariance manifest.

\section{Hamiltonian and Action}

The abelian Born-Infeld action on the D4-brane is given by 
\begin{equation}
S_{BI} = \int\! d^4xdt\, 
            \frac{-1}{\lambda}\sqrt{\det\bigl(-g-\CF\bigr)}
            +\ha c_1\wedge\CF\wedge\CF +c_3\wedge\CF,
\end{equation}
where $\CF=F-b$. Here we use units where the ten-dimensional string 
length $\ell_s=1$. The coupling constant $\lambda$ in front of the first 
term is the $10$ dimensional 
string coupling; the fact that it appears as $1/\lambda$ is typical of an 
RR-soliton. The coupling to the RR potentials $c$ is obtained 
by expanding $c\wedge\exp{\CF}$. 

We want to obtain the Hamiltonian associated to this action, and express 
it in terms of fields whose zero-modes are the various quantum numbers. 
These are
\ba
\mbox{electric fields} &:& E^i = \frac{\delta\CL}{\delta F_{0i}},\nn\\
\mbox{magnetic fields} &:& F_{ij} = \partial_{[i}A_{j]},\nn\\
\mbox{momenta}         &:& P_i = F_{ij}E^j,\\
\mbox{instanton density}&:& \CI = \ha F\wedge F,\nn\\
\mbox{rank}            &:& 1.\nn
\ea
Straightforward calculation gives the result
\begin{eqnarray}
\CH &=& E^i\dot A_{i}-\CL_{BI} \\
  &=& \Biggl[\frac{1}{\lambda^2}\biggl(\det g +\frac{\det g}{2}(\CF_{ij})^2
            +\frac{1}{4}(\CF\wedge\CF)^2\biggr) \nonumber\\
  &&+\begin{pmatrix} E'\\ P' \end{pmatrix}^t
     \begin{pmatrix} g-bg\inv b&bg\inv\\ -g\inv b&g\inv \end{pmatrix}
     \begin{pmatrix} E'\\ P' \end{pmatrix}
   \Biggr]^{1/2}. \nonumber
\end{eqnarray}
The shifted primed electric field and momentum $E'$ and $P'$ are given by
\begin{equation}
E'=E-\st \CF c_1+\st c_3,\qquad
P'=P+\ha \CF\wedge\CF c_1+\CF\,\st c_3.
\end{equation}

We want to rewrite this Hamiltonian such that its U-duality properties 
become manifest. To this end we first have to group together the different 
fields in five-dimensional representations, motivated by their M-theory 
interpretation. The four-momenta $P_i$ sit together in a five-dimensional 
momentum vector with the instanton number density $P_5\equiv\frac12 
F\wedge F$; the latter, indeed, is associated to the D0-branes, which are 
M-theory momentum modes around the extra circle. Similarly, the electric 
fields are associated to winding strings and therefore are joined to the 
membrane winding fields $\st F$, the four-dimensional hodge dual of the 
magnetic field. Together they form an antisymmetric tensor,
\be
M_{ij}=\st F_{ij}, \quad M_{i5}= E_i.
\ee
Finally, the rank $N$, the number of five-branes, is a scalar under the 
$\SO(5)$-rotation group. In the present calculation, $N=1$. In 
non-abelian gauge theory, the action involves a trace over the group 
indices, giving $N$ as $\mbox{Tr }1$. In the following we will for 
clarity use $N$ instead of $1$.

In order to change to the M-theory picture, we also upgrade the moduli to 
five-dimensional ones. The $b$- and $c_3$-fields are components of the 
eleven-dimensional three-form $C$, whereas the coupling $\lambda$ and the 
RR 1-form $c_1$ come from the 11-dimensional metric,
\be
G= \begin{pmatrix}
    g_{ij}+ \lambda^2{c_1}^i{c_1}^j& -\lambda^2{c_1}^i\\
    -\lambda^2{c_1}^j& \lambda^2
\end{pmatrix},
\ee
in string units.

To make the $\SO(5,5)$ U-duality manifest, it is appropriate to convert 
to six-dimensional Planck units, 
\be
\ell_{pl}=\lambda^{\frac12}{(\sqrt{\det{g}}/\ell_s^4)^{-\frac14}}\ell_s 
\ee
(see \ref{planck}), since this is the length scale that is U-invariant.
After a rescaling of the metric with a factor $(\det G)^{-1/3}$, the 
Hamiltonian has the form, in units $\ell_{pl}=1$, 
\be
\label{hamfinal}
\CH^2 = \sqrt{\det G} G^{ij}  
P'_i P'_j
 + \frac{1}{2\sqrt{\det G}} G_{ik} 
G_{jl} {M'}^{ij}{M'}^{kl} + \frac{1}{\sqrt{\det G}} 
N'^2, 
\ee

The three-form field moduli enter in this formula as shifts in the 
fields, indicated by the primes; this dependence on the 
three-form $C$, or equivalently its five-dimensional dual 
two-form $\st C$, can be conveniently written, if we identify the 
vector $P$ with a four-form, as 
\ba
\label{cshift}
P' + M' + N' &=& e^{-\st{C}}\wedge (P + M + N) 
\nn\\
          &=& (P - \st{C}\wedge M + \ha \st{ C}\wedge 
\st{ C} N) + (M - \st{C}N) + N.
\ea

If we assume that we can extend the manifest $\SL(4)$ to an $\SL(5)$ 
action on the fields $P_i$ and $M^{ij}$, the 
Hamiltonian would be manifestly invariant under the 
$\SL(5,\Z)$ coming from the five-torus in M-theory. As we shall see in the 
next 
section, this symmetry can even be extended to the full U-duality group, 
$\SO(5,5,\Z)$.

\section{U-Duality and BPS-Spectrum}

The Hamiltonian that we have derived from the four-dimensional Born-Infeld 
action (\ref{hamfinal}) depends on sixteen fields (not all independent) 
whose zero modes are integer quantum numbers (note that one of these fields 
is the rank of the gauge group; it is hard to imagine whether it has any 
modes other than the zero mode). These sixteen quantum numbers, and hence 
also the fields, sit in the sixteen-dimensional spinor representation of the 
U-duality group, which is a discrete subgroup of $\SO(5,5)$. The 
Hamiltonian (\ref{hamfinal}) is invariant under these U-duality 
transformations. To see how they act it is convenient to represent the 
fields in terms of a bispinor of $\SO(5)$, satisfying a reality condition; 
this is equivalent to a spinor of $\SO(5,5)$.
We set $\ell_{pl}=1$, and introduce the bispinor
\be
Z =  \frac{N'}{(\det G)^{1/4}} + (\det G)^{1/4} P'_i\Gamma^i
+ \frac{1}{2(\det G)^{1/4}} {M'}^{ij}\Gamma_{ij},
\ee
where the $\Gamma$'s are five-dimensional hermitian gamma-matrices satisfying
\be
\{ \Gamma_i, \Gamma_j\} = 2 G_{ij}.
\ee
The Hamiltonian then takes the form of the invariant
\be
\CH = \frac{1}{2}\sqrt{\tr Z Z^\dag} \equiv \| Z \|.
\ee
We now set out to derive a BPS-bound for this system, i.e. a lower bound 
on the mass given the values of the various quantum numbers. To obtain 
this bound we introduce the vector \cite{dvv}
\be
K_L = \frac{1}{8}\mbox{tr}\Gamma ZZ^\dag.
\ee   
This vector $K_L$ can be expressed in terms of the various fields as follows.
It is a linear combination of two vectors $K$ and $W$, which in the
absence of the $\st{C}$ two-form are given by
\be
K_i =  N P_i - \frac{1}{2}(M\wedge M)_i
,\quad W^i=M^{ij}P_j,\quad {K_L}_i = K_i + G_{ij}W^j.
\ee
First of all, one can easily check (using the Schouten identity, see 
\ref{schouten}) that upon inserting the original gauge theory 
expressions, both vectors are identically
zero. If we switch on the $\st{C}$-field, we have to shift $P_i$ and 
$M^{ij}$: 
\ba
P_i&\to& P_i - (\st{C}\wedge M)_i +
\frac{1}{2}(\st{C}\wedge\st{C})_i N \nonumber \\
M &\to & M - \st{C} N .
\ea
Inserting this one immediately finds that $K$ remains unaffected. Using
the Schouten identity in five dimensions (see \ref{schouten}), we 
find \be
W^i\to W^i - \st{C}^{ij}K_j.
\ee
Hence also for general $\st{C}$ the total $K_L$ vanishes. (The vanishing of 
these vectors is precisely equivalent to the fact that the various fields 
are dependent.) 

Using this vector $K_L$,  
we can now derive a BPS-mass formula in terms of the fluxes, which are 
the zero-modes of the bispinor $Z$, denoted as $Z^0$. The details are 
presented in \ref{BPS}. The result is
\be
M_{BPS}^2 =  \| Z^0\|^2 + 2 |K_L^0|.
\ee
$K_L^0$ is defined as the part of $K_L$ originating from the zero-modes 
of $Z$,
\be
K_L^0 = \frac18 \mbox{tr}\Gamma Z^0{Z^0}^\dag.
\ee
In components the mass formula reads as follows: 
\ba
M_{BPS}^2 &=& \sqrt{\det G} G^{-1} {n'}^{2} + \ha \sqrt{\det G}\inv 
G^2 {m'}^{2}
 + \sqrt{\det G}\inv N^2 \nonumber\\
&& + 2\sqrt{ \begin{pmatrix} K^0\\W^0\end{pmatrix}
\begin{pmatrix}G\inv -\st{C}G\st{C} & \st{C}G\\
-G\st{C} & G \end{pmatrix}\begin{pmatrix} 
K^0\\W^0\end{pmatrix}}.
\ea 
The $n',m'$ are the zero-modes of the five-dimensional 
fields, shifted with the terms involving $\st C$, as in equation 
(\ref{cshift}). We recognise in the 
first line the $\SO(5,5)$ invariant constructed from the spinor, in the 
second line the one from the vector.

Finally, note that if we replace the rank $N$ by a dual five-form field
strength, and similarly $m'$ by a three-form, the first line in the mass
formula attains a more natural form, with just one overall volume factor
$\sqrt{\det G}$.

\section{Discussion}

From a matrix theory perspective \cite{bfss}, the theory of $N$ D4-branes 
describes DLCQ M-theory compactified on a four-torus 
\cite{tay}. Therefore, one would expect it to have a duality group 
$\SL(5,\Z)$, the U-duality group in seven dimensions. Indeed, in 
\cite{eli,piol,obers}, the masses of the $1/2$ BPS-states where 
derived, 
and shown to transform in $\SL(5,\Z)$ representations. These states 
correspond in our notation with those having $K^0_L=0$. It was realised 
that this group could be enhanced to $\SO(5,5,\Z)$ in \cite{obers,hacq,  
blau} by adding the generator of Nahm transformations, that exchanges the 
rank of the gauge group with the fluxes. Furthermore, in \cite{hacq} 
degeneracies of 
a subset of the BPS-states of Yang-Mills theory in three space 
dimensions were shown to be symmetric under the relevant U-duality group.    

In this paper we demonstrated that the full Hamiltonian of the D4-brane, and 
its BPS mass formula, are symmetric under the U-duality group 
$\SO(5,5,\Z)$, provided one uses the full Born-Infeld action. 
The BPS-spectrum coincides with the one found in \cite{dvv} from the 
M-theory five-brane, which is natural since the D4-brane is the M5-brane 
wrapped around the eleventh dimension. Just as in that case, we conclude 
that the D4-brane action is capable of reproducing the correct BPS 
spectrum of M-theory compactified to six dimensions; a proviso is that 
the five-brane wrapping number, equal to the rank of the gauge group, be 
non-vanishing for this description to make sense. 

We have seen that the rank $N$ is treated on an equal footing with the 
fluxes. This suggests that 
there should be a description of the theory 
where $N$ is treated as a field, with zero-mode equal to the rank, but 
also with fluctuations, $N'$. 

Finally, the procedure carried out in this paper can be easily 
generalised to dimensions lower than $4+1$, giving the appropriate 
$\E_{d+1}$ U-duality symmetric mass formulae. In higher dimensions extra 
quantum numbers corresponding to e.g. NS5-branes, which are not naively 
present in the gauge theory, 
have to be added.

\ack

The authors have benefited from discussions with Feike Hacquebord, Yolanda 
Lozano, Niels Obers and Boris Pioline. C.H. and G.Z. are supported 
financially by Stichting FOM. The research of E.V. is partly supported by 
the Pionier Programme of the Netherlands Organisation for Scientific 
Research (NWO).

\appendix

\section{The Schouten Identity}
\label{schouten}

The Schouten identity states that any 5-rank antisymmetric tensor in 4 dimensions 
must vanish. It can be written in the form 
\begin{equation}\label{schout}
\ep_{ijkl}\delta_m^n - \ep_{ijkm}\delta_l^n - \ep_{ijml}\delta_k^n 
 - \ep_{imkl}\delta_j^n - \ep_{mjkl}\delta_i^n=\ep_{[ijkl}\delta_{m]}^n=0.
\end{equation}
Contracting this identity with antisymmetric matrices $A_{ij}$ and $B_{kl}$, 
we find the matrix products 
\begin{equation}
{}^*AB+{}^*BA = -A\wedge B\one,
\end{equation}
where 
\begin{equation}
{}^*A_{ij}=\ha\ep_{ijkl}A_{kl},
\qquad\mbox{and}\qquad
A\wedge B = \frac{1}{4}\ep_{ijkl}A_{ij}B_{kl} 
 = \frac{1}{2}{}^*A_{ij}B_{ij}.
\end{equation}

In 5 dimensions, there is a similar relation $\ep_{[ijklm}\delta_{n]}^p=0$. Contracting 
now with three antisymmetric matrices $A$, $B$ and $C$, we find the relation 
\begin{equation}
A_{ij}(B\wedge C)^j + B_{ij}(C\wedge A)^j + C_{ij}(A\wedge B)^j = 0,
\end{equation}
where 
\begin{equation}
(A\wedge B)^i = \frac{1}{4}\ep^{ijklm}A_{jk}B_{lm}.
\end{equation}

\section{Planck Units}
\label{planck}
  
The relation between the 11-dimensional Planck length $\ell_p$ and the string 
length $\ell_s$, is
given by
\begin{equation}
\ell_p=\lambda^{1/3}\ell_s,
\end{equation}
where $\lambda$ is the string coupling. 
The Planck length in $11-d$ dimensions, after compactification on a 
$d$-torus, is given by
the formula
\begin{equation}
\ell_{pl}=\ell_p (V_d)^{-1/(9-d)},
\end{equation}
where $V_d$ is the volume of the internal $d$-torus in 11-dimensional 
Planck units. 
For the case at hand, $d=5$,
we find $\ell_{pl}=\ell_p 
\lambda^{1/6}v^{-1/4}=\lambda^{1/2}v^{-1/4}\ell_s$.
The factor $v$ in this expression is the original, four-dimensional 
volume, $\sqrt{\det g}$, measured in string units $\ell_s$. The squared 
volumes of the five-torus, in the various units, are 
\be
\det{\frac{G}{\ell_s^2}} = v^2 \lambda^2,\quad \det{\frac{G}{\ell_p^2}} = 
\frac{v^2}{\lambda^\frac{4}{3}}, \quad \det{\frac{G}{\ell_{pl}^2}} = 
\frac{v^{\frac{9}{2}}}{\lambda^3}, \ee
(again with $\ell_s^4 v=\sqrt{\det{g}}$).
In our final expression for the Hamiltonian we rescale the metric with a 
factor $\lambda/v^{3/2}$, which 
equals the determinant in six-dimensional Planck units to the power $-1/3$. 
Therefore, the determinant that appears in equation (\ref{hamfinal}) is
\be
\det G = \left(\frac{\lambda}{v^{3/2}}\right)^5 
\frac{v^{\frac{9}{2}}}{\lambda^3} = \frac{\lambda^2}{v^3}.
\ee

\section{The BPS-bound}
\label{BPS}

The BPS-bound is the minimum of the Hamiltonian, 
which is derived using a Bogomolny type argument. We introduced the 
bispinor $Z$, in terms of which the Hamiltonian is defined as
$$
\CH = \frac{1}{2}\sqrt{\tr Z Z^\dag} \equiv \| Z\|.
$$
Furthermore we need the vector $K_L=\frac18 \mbox{tr}\Gamma ZZ^\dag$, 
which is identically zero, as argued in the text. Therefore, we can, 
for any unit five-vector $e$, 
write $\CH^2$ 
as 
\be 
\CH^2 = \frac{1}{2}\| Z+e\Gamma Z \|^2 \equiv \ha \| Z( e)\|^2. 
\ee
The mass of a state is given by the integrated hamiltonian density $\CH$,
\be
M = \ha \sqrt{2}\int \| Z(e)\|.
\ee
Now we split $Z(e)$ into a piece $Z^0(e)$ containing only the 
zero-modes of the fields, which integrate to the flux quantum numbers, and 
the non-zero-modes $Z'(e)$ (that integrate to zero).
Using the inequality
\be
\| Z(e)\| \geq \| Z^0(e)\| + \frac{Z^0(e)\cdot Z'( 
e)}{\| Z^0(e)\| },
\ee
we obtain the mass-bound
\be
M^2 \geq \ha \left( \int \| Z^0(e)\|\right)^2 = \int \| Z^0\|^2 + 2 
e\cdot K_L^0.
\ee
The BPS-bound can then be obtained be maximising this minimum bound, 
which evidently leads to 
\be
M_{BPS}^2 =  \| Z^0 \|^2 + 2 |K_L^0|.
\ee

\end{document}